# A Generalized approach for Approximate Solutions to the N Body Problem

AbuBakr Mehmood, Syed Umer Abbas Shah and Ghulam Shabbir

Faculty of Engineering Sciences,

GIK Institute of Engineering Sciences and Technology

Topi, Swabi, NWFP, Pakistan

Email: shabbir@giki.edu.pk

## Abstract:

An approach is developed to find approximate solutions to the classical Newtonian problem of N bodies. Sets of N gravitating bodies having spherically symmetric mass distributions, small angular velocities (< 1 rad/s) and bounded position vectors have been taken into consideration. In addition, it is assumed that the masses form an isolated system in free space and perform free gravitating motion. Although the problem is not exactly solvable, a new approach will be developed to find approximate solutions using N number of two body motion analogues.

## Introduction:

As mentioned in the abstract, the problem chosen for analysis is not an exactly solvable problem. We will therefore aim at finding approximate solutions and other approaches can be found in [1-5]. Before proceeding with our discussion on the procedure that we have developed for solution, we present a formal definition of the problem.

The problem by definition is to solve explicitly for each of the position vectors of $N$ gravitating masses, all of which perform free motion under each other's gravitational attraction in free space. The masses form an isolated system in space, and hence the motion of the $kth$ mass say, is under the influence of the remaining $N-1$ masses only. Some of the critical assumptions considered while solving the problem have been briefly discussed in the following paragraph.

We specifically assume that all of the gravitating masses posses mass distributions that are spherically symmetric in nature. This would give us the liberty to approximate each of the N masses as a point mass. Note that this assumption should simplify our derivations, and yet, have negligible effect on the accuracy of our results. The reason behind this fact is that celestial bodies (planets and stars) truly posses mass distributions that are almost exactly spherically symmetric in nature. Hence this assumption, apart form simplifications, should also provide us with nearly an exact replication of the real situation. It is worth mentioning here that the procedure that we have developed is not valid for explosions or collisions. We will consider solving the problem for angular velocities that are small, and position vectors that are bounded. A frame of reference will be attached to the centre of mass of this system, and use of Newton's Laws will be made for Mathematically representation. Since it can be shown that the centre of mass of such a system has zero acceleration for all time, the attachment of a frame of reference to this centre of mass along with the use of Newton's laws is justified. Since we want to represent the generalized situation, we will solve for the motion of the *kth* mass. We represent the position vector of this mass by $\mathbf{r}_k$. Figure 1 shows the diagrammatic illustration of this generalized situation.

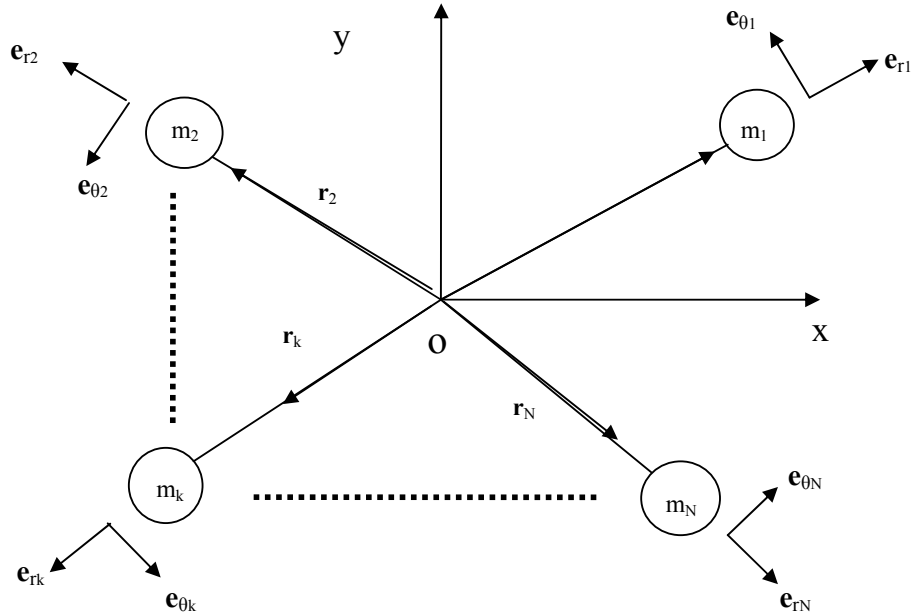

*Figure 1*

Here *oxy* is an inertial frame of reference attached to the centre of mass of the *N* bodies, $\hat{e}_{rk}$ is a radial unit vector along the direction of $\mathbf{r}_k$, $\hat{e}_{\theta k}$ is a unit vector perpendicular to $\hat{e}_{rk}$ (in the direction of increasing rotation angle of $\mathbf{r}_k$) and $\theta_k$ is the rotation angle of vector $\mathbf{r}_k$. Note that we are representing the generalized situation, and the allowable values of $k$ are $1,2,3,...,N$. Here the index $k$ is being used as a label for the masses. When $k=1$, for example, we are representing all associated quantities for the mass $m_1$. We now introduce a few more assumptions. We specifically assume that all the position vectors involved remain bounded i.e. they are not arbitrarily large, and that all the associated angular velocities ($\dot{\theta}_k$, $k=1,2,3,...N$) are considerably smaller than one radians per second. The reason behind these assumptions should become obvious when the solution procedure is presented, since they will serve to considerably simplify our derivations. Commenting on the validity of these added assumptions, we claim that they are practically feasible in the sense that they provide a reasonably accurate replication of the exact situation. Angular velocities involved in celestial orbital motion of planets and stars are considerably smaller than one radians per second. Therefore our assumption of angular velocities being considerably smaller than 1 rad/s is reasonable and should cause negligible inaccuracies in our results. Also, we have considered solving the problem for the case when all the position vectors remain bounded. It would make no sense trying to include the case when any number of position vectors approach unboundedness, since the associated masses would then be essentially free of the gravitational pull of the remaining bodies. The Mathematical translation of this condition would then be $|\mathbf{r}_k| < \infty$. Since we are free to scale the position vectors according to our convenience, it follows that we can always scale them in a manner such that they attain values much smaller than infinity. This gives us the liberty to modify our condition about boundedness of position vectors as $|\mathbf{r}_k| << \infty$. Stated Mathematically, we have assumed the following.

$$|\mathbf{r}_k| << \infty \text{ and } \left|\dot{\theta}_k\right| << 1 rad/s \; \forall \; k = 1, 2, 3, ...N \tag{1}$$

We now introduce another critical assumption. We assume that while executing free motion with the $N-1$ bodies, the *kth* body having mass $m_k$ and position vector $\mathbf{r}_k$ remains approximately in two body motion with a body of mass $M_k = \sum_{n=1}^{N} m_n$ ($n \neq k$) placed at a point given by the centre of mass of the remaining $N-1$ bodies,

$$\mathbf{r}_{M_k} = \frac{\sum_{n=1}^{N} m_n \mathbf{r}_n \ (n \neq k)}{\sum_{n=1}^{N} m_n \ (n \neq k)}.$$ Figure 2 presents a diagrammatic illustration of this situation.

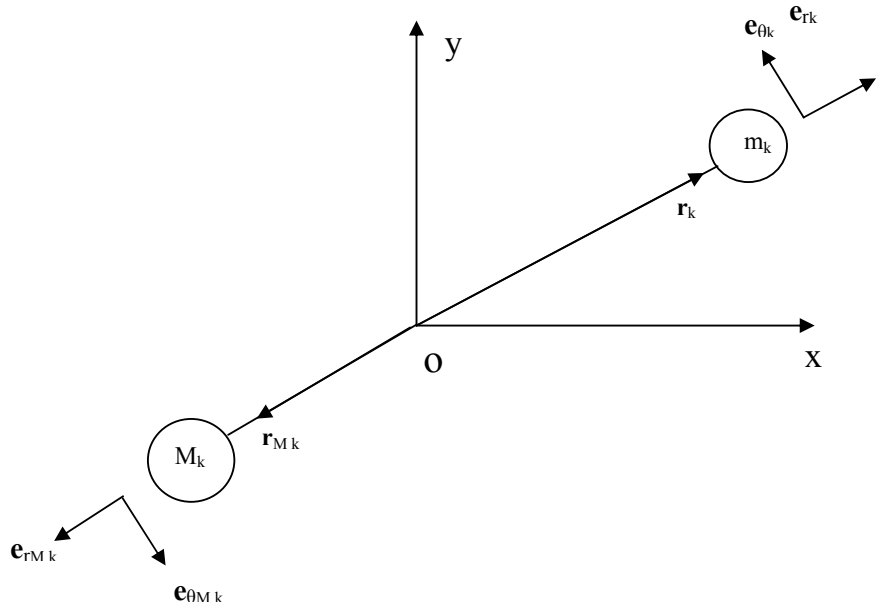

*Figure 2*

Here $\mathbf{r}_k$ and $\mathbf{r}_{M_k}$ are the position vectors of bodies $m_k$ and $M_k$ respectively. Also $\hat{e}_{r_k}$ is a radial unit vector in the direction of $\mathbf{r}_k$, and $\hat{e}_{\theta_k}$ is a unit vector perpendicular to $\hat{e}_{r_k}$, in the direction of increasing $\theta_k$. $\theta_k$ is the rotation angle for vector $\mathbf{r}_k$. Similar arguments hold for $\hat{e}_{r_{Mk}}$ and $\hat{e}_{\theta_{Mk}}$ in the case of vector $\mathbf{r}_{M_k}$. The position vector of $m_k$ and $M_k$ i.e. $\mathbf{r}_k$ and $\mathbf{r}_{M_k}$ respectively, are assumed to remain approximately collinear for all time since the motion between $m_k$ and $M_k$ has been assumed to be approximately two body motion. It then follows that the respective unit vectors in those directions namely $\hat{e}_{r_k}$ and $\hat{e}_{r_{Mk}}$ also remain approximately collinear for all time i.e. $\hat{e}_{r_k} \cdot \hat{e}_{r_{Mk}} = -1 \ \forall \ t$. Note that by using this assumption we have in effect considered replicating $N$ body motion by the use of $N$ number of two body motion analogues, one for each body in turn. We claim that this assumption is only approximately true, had it been exactly true, the $N$ body problem would have been reducible to the two-body problem, and exact solutions could have been obtainable. Having had an adequate introductory discussion, we now go on to present the formal procedure for our solutions.

## Main Results:

We will quite often need to express the position vectors as time functions being multiplied by the respective radial unit vectors. For this purpose we will use the notation $\mathbf{r}_k(t) = r_k(t)\hat{e}_{r_k}$ and $\mathbf{r}_{M_k}(t) = r_{M_k}(t)\hat{e}_{r_{Mk}}$. Now we define the vector $\mathbf{x}_k = \mathbf{r}_k - \mathbf{r}_{M_k} = x_k(t)\hat{e}_{xk}$, illustrated in figure 3. Here $\hat{e}_{xk}$ is a radial unit vector in the direction of $\mathbf{x}_k$, and $\hat{e}_{\theta_{xk}}$ is a unit vector perpendicular to $\hat{e}_{xk}$ in the direction of increasing $\theta_{xk}$. $\theta_{xk}$ is the rotation angle of vector $\mathbf{x}_k$.

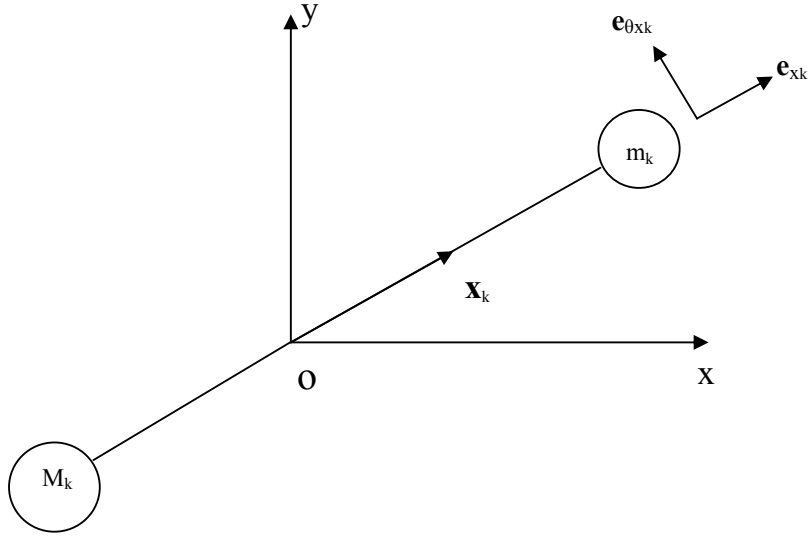

*Figure 3*

Making use of the figures 2 and 3, the following relations can be shown to hold true.

$$\mathbf{x}_k = \mathbf{r}_k - \mathbf{r}_{M_k} \tag{2a}$$

$$\hat{e}_{xk} = \hat{e}_{r_k} = -\hat{e}_{r_{Mk}} \tag{2b}$$

$$\hat{e}_{\theta_{xk}} = \hat{e}_{\theta_k} = -\hat{e}_{\theta_{Mk}} \tag{2c}$$

$$x_k(t) = r_k(t) + r_{M_k}(t) \tag{2d}$$

$$\theta_{xk}(t) = \theta_k(t) = \theta_{M_k}(t) + \pi \tag{2e}$$

$$\dot{\theta}_{xk}(t) = \dot{\theta}_k(t) = \dot{\theta}_{Mk}(t) \tag{2f}$$

Modeling the system in figure 2 now by the use of Newton's second law and

Newton's law for gravitation, and making use of $(2b)$ and $(2d)$, we get

$$\ddot{\mathbf{r}}_k = -\left(\frac{GM_k}{x_k^2}\right)\hat{e}_{r_{xk}} \tag{3a}$$

$$\ddot{\mathbf{r}}_{Mk} = \left(\frac{Gm_k}{x_k^2}\right)\hat{e}_{x_k} \tag{3b}$$

We could now make use of $(2a)$ to express this model in a more compact form

$$\ddot{\mathbf{x}}_k = -\left(\frac{G(m_k + M_k)}{x_k^2}\right)\hat{e}_{x_k} \tag{3c}$$

Resolving $(3c)$ in polar coordinates and comparing coefficients of the unit vectors on both sides of the equation, we can obtain the following scalar analogue of $(3c)$.

$$\ddot{x}_k - x_k \dot{\theta}_k^2 = -\left(\frac{G(m_k + M_k)}{x_k^2}\right) \tag{4a}$$

$$x_k \ddot{\theta}_k + 2\dot{x}_k \dot{\theta}_k = 0 \tag{4b}$$

We argued at the beginning that the angular velocities involved were smaller than 1 radians per second, and that the position vectors were not arbitrarily large. In (1), we have already provided a mathematical representation of this fact.

Considering the expression defining $\mathbf{r}_{M_k}$ and taking into account (1), we can conclude that $r_{M_k}(t) << \pm\infty$. Taking this fact and $(2d)$ into consideration, it can be shown that $x_k(t) << \pm\infty$. Use of this information along with $(2e)$ allows us to set $x_k \dot{\theta}_k^2 \simeq 0$ in equation $(4a)$, which then takes the form

$$\ddot{x}_k = -\left(\frac{G(m_k + M_k)}{x_k^2}\right) \tag{4c}$$

making use of $(4b)$ we can show that

$$x_k^2 \dot{\theta}_k = x_{ko}^2 \dot{\theta}_{ko} \tag{4d}$$

where $x_{ko} = r_{ko} = r_k(0)$, and $\dot{\theta}_{xko} = \dot{\theta}_{ko} = \dot{\theta}_k(0)$. As a next step, multiplication of $(4c)$ by $\dot{x}_k\, dt$ allows us to integrate the resulting equation on the left hand side w.r.t. '$t$' and on the right hand side w.r.t. '$r$', and simplification yields the form

$$\frac{dx_k}{dt} = \pm\left[\frac{A_k}{x_k} + B_k\right]^{\frac{1}{2}} \tag{5a}$$

where $A_k = 2G(m_k + M_k)$ and $B_k = \dot{x}_{ko}^2 - \frac{2G(m_k + M_k)}{x_{ko}}$. We now separate variables and apply integration to both sides of the equation to get

$$\int_{x_{ko}}^{x_k}\left[\frac{A_k}{B_k x_k} + 1\right]^{-\frac{1}{2}} dx_k = \pm\sqrt{B_k}\int_{t_o}^{t} dt \tag{5b}$$

Note that $B_k \geq 0$ in the above relation. Now although (5b) can be integrated in its current form and an implicit equation relating $x_k$ and $t$ can be found, however $x_k(t)$ cannot be explicitly solved for. This fact encourages us to try a simple binomial approximation of the form $\left[\left(\frac{A_k}{B_k x_k}\right) + 1\right]^{-\frac{1}{2}} \simeq 1 - \frac{A_k}{2B_k x_k}$ $\forall$ $|x_k| > |\frac{A_k}{B_k}|$. Use of this approximation simplifies (5b) so that we are capable of integrating on both sides and deriving the following implicit equation

$$x_k + \ln x_k^{-h_k} = f_k(t) \tag{6}$$

where $h_k = \left(\frac{A_k}{2B_k}\right)$ and $f_k(t) = \pm\sqrt{B_k}(t - t_o) + x_{ko} - \ln x_{ko}^{h_k}$. Here again, it should be noted that $B_k$ should not be allowed to attain negative values for the validity of result (6). Solving for $x_k(t)$ explicitly from (6) we get

$$x_k(t) = -h_k * lambertw\left[\frac{-e^{-\frac{f_k}{h_k}}}{h_k}\right] \tag{7}$$

where '$lambertw$' is the notation used for the Lambert's wave function. Replacing the expressions for $h_k$ and $f_k(t)$ into (7) and performing a few manipulations, we can derive the expression

$$x_k(t) = -\left(\frac{A_k}{2B_k}\right) lambertw\left[c_{4k} e^{c_{5k} t}\right] \tag{8}$$

where $c_{4k} = c_{1k} e^{c_{2k} t_o}$, $c_{5k} = -c_{2k}$, $c_{1k} = -\left(\frac{2B_k}{A_k}\right) e^{(-\frac{2B_k}{A_k})(x_{ko} - \ln x_{ko}^{\frac{A_k}{2B_k}})}$ and $c_{2k} = \pm\left(\frac{2B_k\sqrt{B_k}}{A_k}\right)$.

Having computed $x_k(t)$, we will now make use of relation (3a) to find $r_k(t)$ and $\theta_k(t)$. It is worth stating directly that in a manner similar to one adopted in the derivation of $(4c)$ and $(4d)$ from $(3c)$, we can derive the following set of relations from $(3c)$.

$$\ddot{r}_k = -\left(\frac{GM_k}{x_k^2(t)}\right) \tag{9a}$$

$$r_k^2 \dot{\theta}_k = r_{ko}^2 \dot{\theta}_{ko} \tag{9b}$$

We now substitute (8) in (9a) and integrate twice w.r.t. to get our approximation for $r_k(t)$.

$$r_k(t) = h_{1k} + \left(\frac{h_{2k}}{c_{5k}}\right)\left[\frac{1}{2}(lambertw[c_{4k}e^{c_{5k}t}])^4 + (lambertw[c_{4k}e^{c_{5k}t}])^3 + \frac{1}{2}(lambertw[c_{4k}e^{c_{5k}t}])^2\right]$$

$$-h_{1k}t_o - \left(\frac{h_{2k}}{c_{5k}}\right)\left[\begin{array}{c}\frac{1}{2}(lambertw[c_{4k}e^{c_{5k}t_o}])^4 + \\ (lambertw[c_{4k}e^{c_{5k}t_o}])^3 + \frac{1}{2}(lambertw[c_{4k}e^{c_{5k}t_o}])^2\end{array}\right] + r_{ko} \tag{10}$$

where $r_{ko} = r_k(0)$, $\dot{r}_{ko} = \dot{r}_k(0)$, $h_{1k} = \dot{r}_{ko} - \left(\frac{h_{ak}}{2c_{5k}}\right)\left[1 + 2lambertw(c_{4k}e^{c_{5k}t_o})\right]$,

$h_{2k} = \left(\frac{h_{ak}}{2c_{5k}}\right)$ and $h_{ak} = -\left(\frac{4B_k^2 GM_k}{A_k^2}\right)$. Having accomplished the above result, we are now free to substitute equation (10) in (9b), separate variables, and integrate w.r.t. time to obtain the following explicit solution for $\theta_k(t)$.

$$\theta_k(t) = \theta_{ko} + \frac{4x_{ko}^2 \dot{\theta}_{ko} B_k^2}{c_{5k}A_k^2}(1 + 2lambertw[c_{4k}e^{c_{5k}t_o}])(lambertw[c_{4k}e^{c_{5k}t_o}])^2$$

$$-\left(\frac{4x_{ko}^2 \dot{\theta}_{ko} B_k^2}{c_{5k}A_k^2}\right)(1 + 2lambertw[c_{4k}e^{c_{5k}t}])(lambertw[c_{4k}e^{c_{5k}t}])^2 \tag{11}$$

It should be noted that we have successfully approximated the motion of body $m_k$, given by (10) and (11). We now go on to sum up the conditions for which this solution is valid. First of all, we require that (5) and (6) hold true. Secondly this solution is rendered infeasible in case of collisions or explosions. Recall that the condition required for the validity of the binomial approximation used to simplify (5*b*) was $|\mathbf{x}_k| > \left|\dfrac{A_k}{B_k}\right|$. Also, a careful look at the various equations encountered while solving the problem, should help us to conclude that we also require $B_k$ to be positive. Reason being that the explicit solution (6) would not hold true, in case of any complex terms arising in the definition of $f(t)$ (defined below (6)). Also, since $B_k$ occurs in various denominators, it cannot be allowed a zero value. Therefore we require $B_k > 0$. As a concluding remark we state that by solving for $r_k(t)$ and $\theta_k(t)$, we have in effect described the vector $\mathbf{r}_k(t)$ and thus completely defined the motion of the *kth* body having mass $m_k$.

## Summary:

In this paper, we developed an approach to solving the classical Newtonian problem of N bodies. It was assumed that the bodies have spherically symmetric mass distributions, small angular velocities (< 1 rad/s) and bounded position vectors. Approximate solutions were then derived using N two body motion analogues to approximately represent the free gravitating motions of N bodies. The Poincare's Dictum comprehensively proves that the problem is not exactly solvable, and is doubtlessly one of the oldest of unsolved problems of classical mechanics. Applications of the problem could vary from describing the evolution of the universe to space mission design.